\documentclass[letter,12pt,showkeys]{revtex4-1}  

\usepackage{hyperref}
\usepackage{xcolor}
\usepackage{graphicx}  
\usepackage{dcolumn}   
\usepackage{bm}        
\usepackage{amssymb}   
\usepackage{gensymb}
\usepackage{color}  
\usepackage{natbib}

\hypersetup{
    colorlinks,
    linkcolor={blue!50!black},
    citecolor={blue!50!black},
    urlcolor={blue!80!black}
}

\begin{document}

\title{Optimizing Dirac fermions quasi-confinement by potential smoothness engineering}

\author{B. Brun$^{1}$, N. Moreau$^1$, S. Somanchi$^2$, V.-H. Nguyen$^1$, A. Mre\ifmmode \acute{n}\else \'{n}\fi{}ca-Kolasi\ifmmode \acute{n}\else \'{n}\fi{}ska$^3$,\\
K. Watanabe$^4$, T. Taniguchi$^4$, J.-C. Charlier$^1$, C. Stampfer$^2$ \& B. Hackens$^1$}
\affiliation{$^1$IMCN/NAPS \& MODL, Universit\'e catholique de Louvain (UCLouvain), B-1348 Louvain-la-Neuve, Belgium }
\affiliation{ $^2$ JARA-FIT and 2nd Institute of Physics - RWTH Aachen, Germany }
\affiliation{$^3$ AGH University of Science and Technology - Krak\'ow, Poland}
\affiliation{$^4$ National Institute for Materials Science, Namiki, Tsukuba, Japan}

\begin{abstract}
With the advent of high mobility encapsulated graphene devices, new electronic components ruled by Dirac fermions optics have been envisioned and realized.
The main building blocks of electron-optics devices are gate-defined p-n junctions, which guide, transmit and refract graphene charge carriers, just like prisms and lenses in optics. 
The reflection and transmission are governed by the p-n junction smoothness, a parameter difficult to tune in conventional devices.
Here we create p-n junctions in graphene, using the polarized tip of a scanning gate microscope, yielding Fabry-P\'erot interference fringes in the device resistance. 
We control the p-n junctions smoothness using the tip-to-graphene distance, and show increased interference contrast using smoother potential barriers.
Extensive tight-binding simulation reveal that smooth potential barriers induce a pronounced quasi-confinement of Dirac fermions below the tip, 
yielding enhanced interference contrast. On the opposite, sharp barriers are excellent Dirac fermions transmitters and lead to poorly contrasted interferences.
Our work emphasizes the importance of junction smoothness for relativistic electron optics devices engineering. 
\end{abstract}


\date{\today}
\email{boris.brun@uclouvain.be, benoit.hackens@uclouvain.be}

\maketitle

In semiconductor technology, the charge carriers density profile governs the devices' properties.
The so-called space charge zone is of fundamental importance in diodes, transistors or solar cells, and its control at the microscopic scale is a prerequisite to reach the desired properties.
In graphene, a semi-metal hosting massless Dirac fermions \cite{Novoselov-2005}, the density profile of a p-n junction plays a really peculiar role.
Provided that electronic transport is ballistic, the ratio between the junction width and the Fermi wavelength governs the transmission and refraction
properties of charge carriers. 
In particular, the relativistic Dirac fermions experience Klein tunneling when impinging perpendicularly on a p-n interface \cite{Klein-1929}, 
which ensures them a perfect unitary transmission independent of the potential barrier height \cite{Allain-2011}.
Additionally, a diverging flow of Dirac fermions is refocused at a p-n interface, similarly to photons entering a negative refraction index medium \cite{Cheianov-2007, Milovanovic2015}, 
an effect denoted as Veselago lensing \cite{Veselago-1968}.

These exotic properties of graphene Dirac fermions led a plethora of electron-optics proposals and realizations, such as electronic 
optical fibers \cite{Beenakker-2009,Hartmann-2010,Williams-2011,Rickhaus-2015}, lenses\cite{Cserti-2007,Mu-2011,Agrawal-2014, Wu-2014, Logermann-2015, Lu-2018,Zhang-2018} 
and their advanced design to create highly focused electron beams \cite{Ming-Hao-2017}, and even the combination of different optical elements to create a 
scanning Dirac fermions microscope \cite{Boggild-2017}.
Aside guiding, the partial reflection encountered at p-n interfaces has been proposed in the early days of graphene to create Fabry-P\'erot interferometers with graphene n-p-n junctions \cite{Shytov-2008}.
These interferences have since then been observed in monolayer \cite{Young-2009,Velasco-2009,Nam-2011,Rickhaus-2013,Oksanen-2014,Handschin-2017,Veyrat-2019} as well as multilayer 
graphene \cite{Varlet-2014,Campos-2012}.
In view of potential applications, complex n-p-n junction geometries that fully take advantage of these Fabry-P\'erot interferences 
have already proven useful to build otherwise inaccessible graphene devices, such as reflectors \cite{Morikawa-2017,Graef-2019} and even transistors \cite{Wang-2019}.

A Fabry-P\'erot interferometer consists in two mirrors facing each other, and the transmission probabilities of these mirrors govern the interference fringes contrast.
In graphene, the mirrors are materialized by two p-n junctions, and their transmission properties could in principle be tuned by controlling the p-n junctions width.
However, p-n junctions in graphene are most often created by means of metallic or graphite gates, whose distance to the graphene plane is by essence fixed, so that 
the p-n junction width is fixed by the sample geometry.
Here we use the polarized tip of a Scanning Gate Microscope (SGM) to induce a n-p-n junction, and take advantage of the SGM flexibility to control and characterize the p-n junctions 
width, independently of the potential barriers height.

Scanning gate microscopy (SGM) consists in scanning an electrically polarized metallic tip, acting as a local gate above a device's surface, and mapping out tip-induced device's 
conductance changes \cite{Eriksson-1996}. 
Initially developed to investigate transport in III-V semiconductor heterostructures \cite{Topinka-2001,Jura-2009, Kozikov-2013, Brun-2014},
SGM brought spatially-resolved insights into transport phenomena occurring in graphene devices, through experiments, simulations and their combination
\cite{Schnez-2010,Pascher-2012,Garcia-2013,Cabosart-2017,Bhandari-2016,Shaohua-2016,Mrenca-2015,Mrenca-2016,Mrenca-2017,Petrovic_2017,Dou-2018}.
Recently, we demonstrated the viability of SGM to study ballistic transport in clean encapsulated graphene devices, and reported optical-like behavior of Dirac fermions
using the tip-induced potential as a Veselago lens\cite{Brun-PrB-2019}.

In the present paper, we show that the transmission probabilities of the p-n junctions can be controlled by tuning the SGM tip-to-sample distance.
Analyzing our experimental findings in the light of tight-binding simulations, we show that the interferences contrast results from the Dirac fermions confinement efficiency,
which is governed by the smoothness of the p-n interfaces.

\begin{figure*}[h!]
\includegraphics[width = 0.8\linewidth]{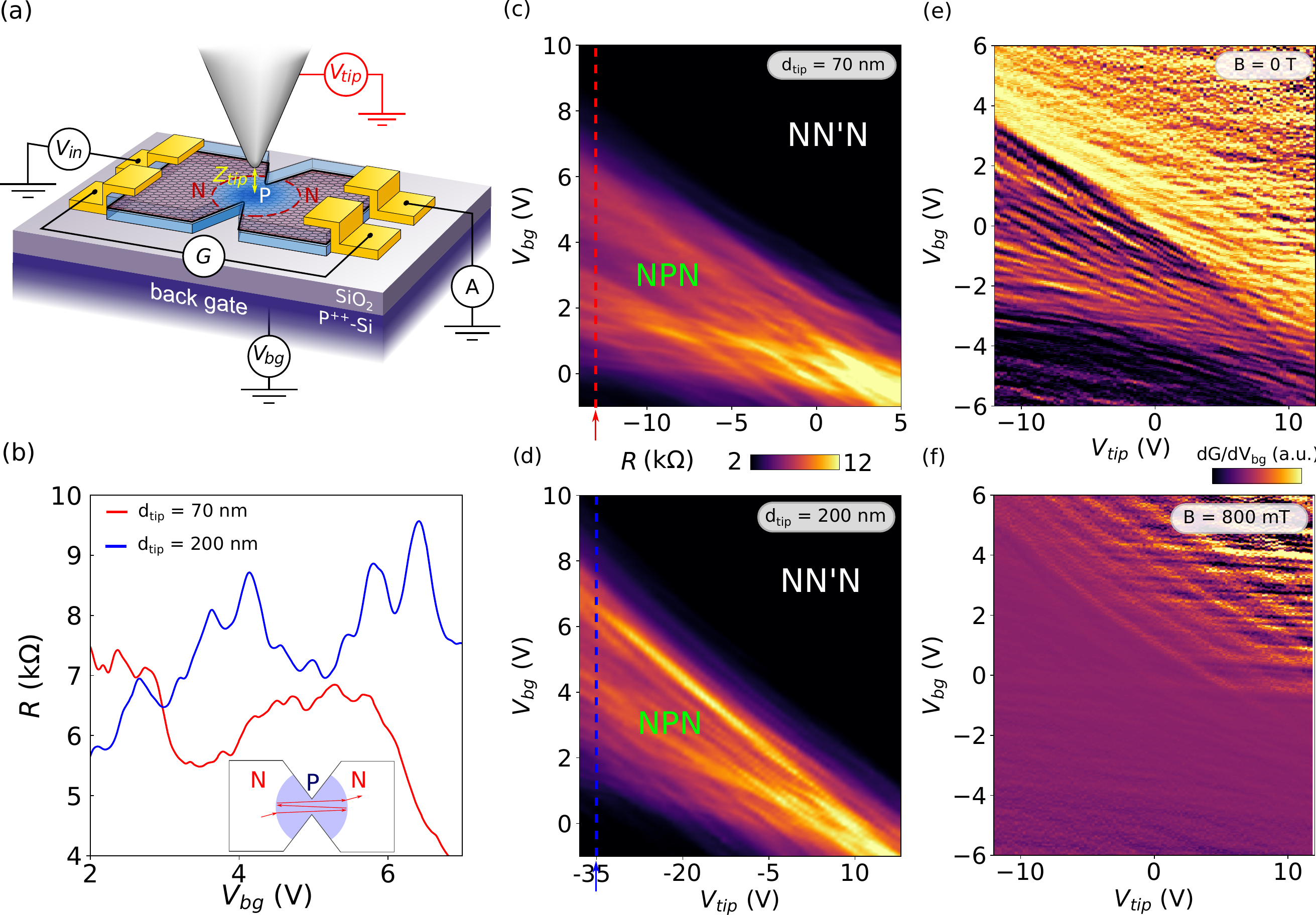}
\caption{\label{fig1} \textbf{Fabry-P\'erot interference:}
(a) Scheme of the experiment: a biased AFM tip is placed above an encapsulated graphene constriction, creating a n-p-n or p-n-p junction. (b) Constriction resistance $R (V_{bg})$ 
for a tip-to-graphene distance $d_{tip}$ = 70 nm (red curve, $V_{tip}$ = -13 V); and $d_{tip}$ = 200 nm (blue curve, $V_{tip}$ = -35 V). In both cases the tip is placed above the constriction center.
(c) Resistance as a function of $V_{tip}$ and $V_{bg}$, for $d_{tip}$ = 70 nm, the tip being placed at the constriction center.(d) 
Resistance as a function of $V_{tip}$ and $V_{bg}$, for $d_{tip}$ = 200 nm. (e) Derivative of conductance versus $V_{tip}$ and $V_{bg}$,
recorded during a different cooldown and with a different tip, and $d_{tip}$ = 100 nm. (f) Same configuration and cooldown as in (e), at a perpendicular magnetic field of 800 mT.}
\end{figure*}

The studied sample is based on graphene encapsulated between two 20~nm-thick hBN layers, in which a 250~nm-wide constriction is defined by etching \cite{Terres-2016}. 
The hBN/graphene/hBN stack lies on top of a highly doped Si substrate covered by a 300~nm $\rm SiO_2$ insulating layer.
This device is thermally anchored to the mixing chamber of a dilution refrigerator, in front of a cryogenic scanning probe microscope \cite{Hackens-2010}.
The device conductance $G$, or resistance $R$, is measured in 4-contacts configuration, by driving a 1~nA current at a frequency of 77,7~Hz, and recording the voltage between two opposite 
contacts using standard lock-in technique (as sketched in Fig.~\ref{fig1}a).
All the data presented here are recorded at a temperature of 100~mK, but global features were found almost independent of temperature up to 1~K,
and even a temperature of 4~K did not noticeably change the observed behavior.
Most of the data presented here were recorded during a single cooldown (except Fig.\ref{fig1}e-f), but this sample showed qualitatively similar behavior for 7 cooldowns.

\begin{figure*}[h!]
\includegraphics[width = 1.0 \textwidth]{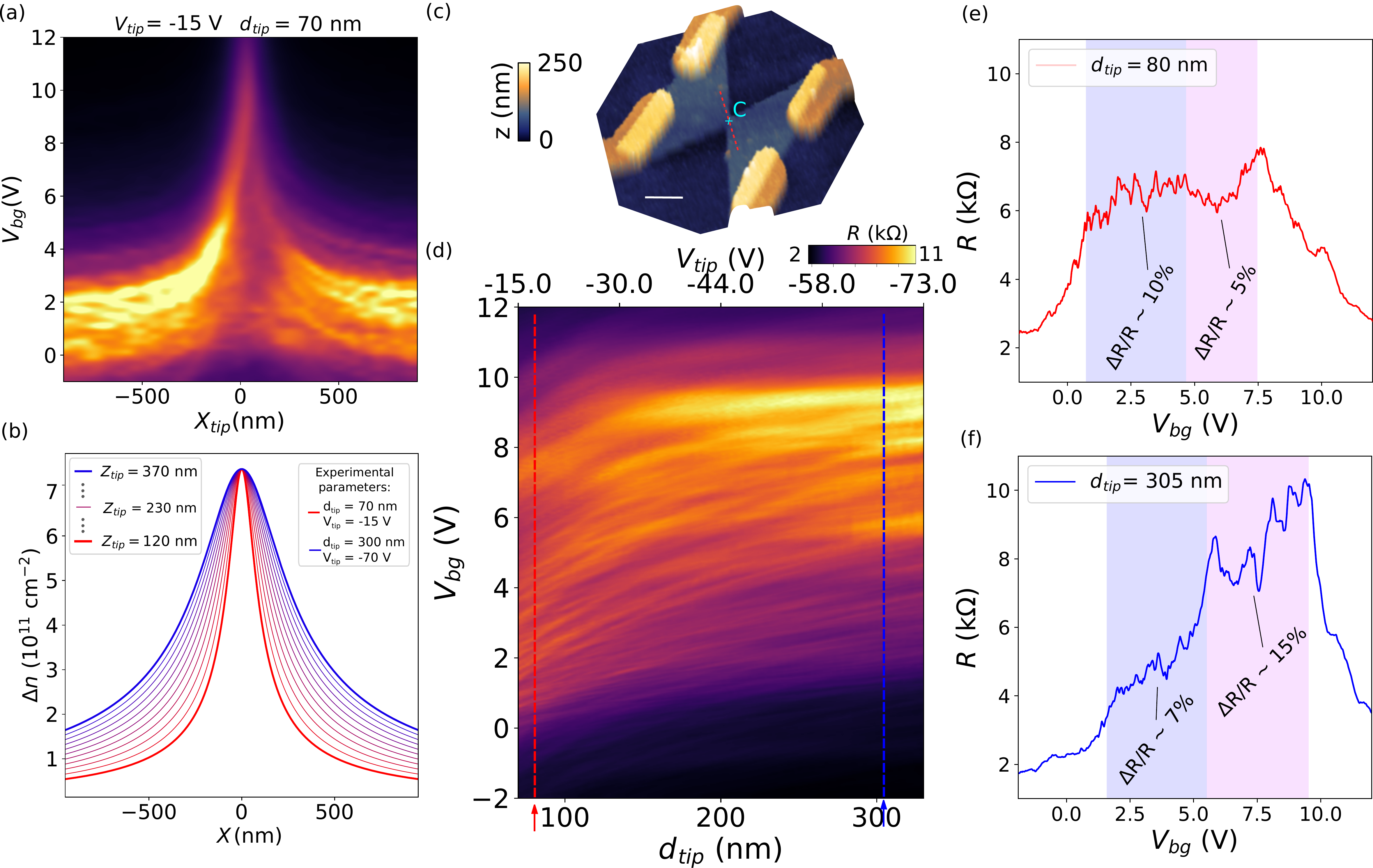}
\caption{\label{fig2} \textbf{Controlling the junction smoothness:}
(a) Resistance as a function of tip position along the blue dashed line in Fig. \ref{fig2}c ($x_{tip} = 0$ being the center of the constriction), and $V_{bg}$. $d_{tip} = 70$ nm, $V_{tip} = -15$ V. 
(b) Estimates of the tip-induced density change as a function of horizontal distance to the tip position ($x$), for different couples of $d_{tip}$ and $V_{tip}$, 
chosen to keep the maximum density change $\Delta n^{max}$ = $7,5.10^{11} \rm cm^{-2}$ while changing the potential extension $R_{tip}$. 
(c) AFM image of the device. Red dashed line represents the line scan used to evaluate the tip potential in (a) and (b).
(d) Resistance as a function of $V_{bg}$ and $d_{tip}$ (lower axis) and $V_{tip}$ (upper axis). The tip is placed at the constriction center, both $d_{tip}$ and $V_{tip}$ are varied to keep 
a constant $\Delta n^{max}$ = $7,5.10^{11} \rm cm^{-2}$ while varying $R_{tip}$. (e,f): Resistance as a function of $V_{bg}$ for two different couples of $d_{tip}$ and $V_{tip}$, extracted from the 
colorplot in Fig.~\ref{fig2}d. (e) $d_{tip}$ = 70 nm and $V_{tip}$ = -18 V. (f) $d_{tip}$ = 305 nm and $V_{tip}$ = -67 V.}

\end{figure*}

The biased SGM tip locally changes the carrier density $n$, leading to a Lorentzian evolution of $n$, centered at the tip position. 
When placing the tip at the center of the constriction, a n-p-n or p-n-p configuration can be reached, depending on the tip voltage $V_{tip}$ and backgate voltage $V_{bg}$.
This is illustrated in Figure~\ref{fig1}c showing resistance as a function of $V_{bg}$ and $V_{tip}$ for a tip-to graphene distance $d_{tip}$ = 70 nm,
The n-p-n region, located at the lower left part of Fig.~\ref{fig1}c, is decorated with a complex pattern of interleaved fringes, resulting from different types of interference phenomena.
In the investigated geometry, one can indeed anticipate that, beside the tip-induced n-p-n or p-n-p junction, other confinements play a 
role and contribute to interferences in the map shown in Fig.~\ref{fig1}c, such as the constriction defined by etching.
Fortunately, increasing the tip-graphene distance to $d_{tip}$ = 200~nm yields a clearer picture, shown in Fig.~\ref{fig1}d, with a much simpler fringe pattern (most of them essentially parallel 
to the n-p-n/n-n'-n limit). The visibility of the pattern is also enhanced by their stronger contrast, when compared to the pattern in Fig.~\ref{fig1}c. 
In Figure~\ref{fig1}b, we plot two profiles of resistance versus $V_{bg}$, for $d_{tip}$ = 70 nm (red curve) and $d_{tip}$ = 200 nm (blue curve) where $V_{tip}$ is adapted 
to reach comparable tip-induced density change (respectively -13 V and -35 V). From this figure, the contrast of the oscillations appears clearly higher for a larger $d_{tip}$,
and a detailed discussion of the origin of this contrast enhancement is one of the main focus of this paper.

It shall first be clarified that these oscillations correspond indeed to Fabry-P\'erot interferences arising inside the tip-induced n-p-n region. 
Figures~\ref{fig1}e and ~\ref{fig1}f (recorded during a different cooldown) illustrate the sensitivity of these interference fringes to a perpendicular magnetic field.
Figure~\ref{fig1}e presents the interference pattern recorded by placing the tip above the constriction center ($d_{tip}$ = 100 nm). 
The map in Fig.~\ref{fig1}e displays the derivative of $G$ versus $V_{bg}$ to highlight the interference fringes, that appear similar to the ones observed in Fig.~\ref{fig1}c-d.
Fig.~\ref{fig1}f shows that they have completely disappeared at a perpendicular magnetic field of 800 mT.
From their characteristic decay field, one can infer that these fringes can be associated with a characteristic length, corresponding to a few hundreds nanometer-long cavity, 
compatible with the cavity formed in the tip induced n-p-n region, as sketched in the inset of Fig.~\ref{fig1}b  (see supplementary data for additional data and a more detailed discussion).
In addition, an accurate determination of the tip-induced potential and an analytical calculation yielding the expected resonances positions in this potential profile agree well with the observed 
oscillations evolution, as detailed below. All these considerations provide strong evidence that the oscillations correspond to Fabry-P\'erot oscillation in the tip-induced n-p-n region.
In the remainder of this paper, we will discuss these interference fringes (Fig.~\ref{fig1}c-e) and show that their visibility depends on the smoothness of the p-n junction, controlled by $d_{tip}$.

As a first step, one needs to precisely evaluate the tip-induced potential. This is done by scanning the tip along the blue dashed line Fig.~\ref{fig2}c while varying $V_{bg}$, at fixed 
$V_{tip}$ and tip-to-graphene distance $d_{tip}$ (\textit{i.e.} the same procedure described in ref. \cite{Brun-PrB-2019}). 
The resulting conductance map shown in Fig.~\ref{fig2}a exhibits a resistance maximum that follows a Lorentzian shape, as the tip crosses the center of the constriction.
This shape is directly related to the shape of the tip-induced potential, as it corresponds to the tip-induced change in the energy of the charge neutrality point 
at the location of the constriction, which governs the device resistance.
Repeating this experiment for several values of $d_{tip}$, and adapting $V_{tip}$ to keep a constant maximum density change below the tip $\Delta n^{max}$, we can fit the
different density profiles under the tip influence, provided that the $V_{bg}$-axis is properly scaled to a density using the backgate lever-arm parameter (see supplementary data).

Considering the tip as a point charge, the expected tip-induced density change would write: $\Delta n(x) = \Delta n^{max}/(1+x^2/Z_{tip}^2)$, where $x$ is the horizontal
distance to the tip center, and $Z_{tip}$ is the effective tip-to-graphene distance, i.e. $Z_{tip}$ = $d_{tip} + a$, $a$ being the tip radius (a = 50~nm).
We define $R_{tip}$ as the half-width at half-maximum (HWHM) of this density profile, that is in this expression given by the effective tip height $Z_{tip}$.
Note that this textbook model underestimates the long-range tail of the tip-induced density change (see supplementary data).
Accurately modeling the tip-induced potential yields a complex electrostatic problem \cite{Zhang-2008,Zebrowski-2018,Chaves-2019}, which is beyond the scope of the present paper.
In turn, we model $\Delta n$ with the following phenomenological equation: $\Delta n(x) = \Delta n^{max}/(\sqrt{1+3x^2/Z_{tip}^2})$, where we assume that the HWHM $R_{tip}$ is given 
by $Z_{tip}$ and is therefore known in the experiment, the only free parameter being $\Delta n^{max}$.
Fig.~\ref{fig2}b shows estimates of tip-induced density changes, for different couples of $V_{tip}$ and $d_{tip}$ leading to the same $\Delta n^{max}$
(see supplementary data and movie for details).

\begin{figure*}[h!]
\includegraphics[width = 1.0 \linewidth]{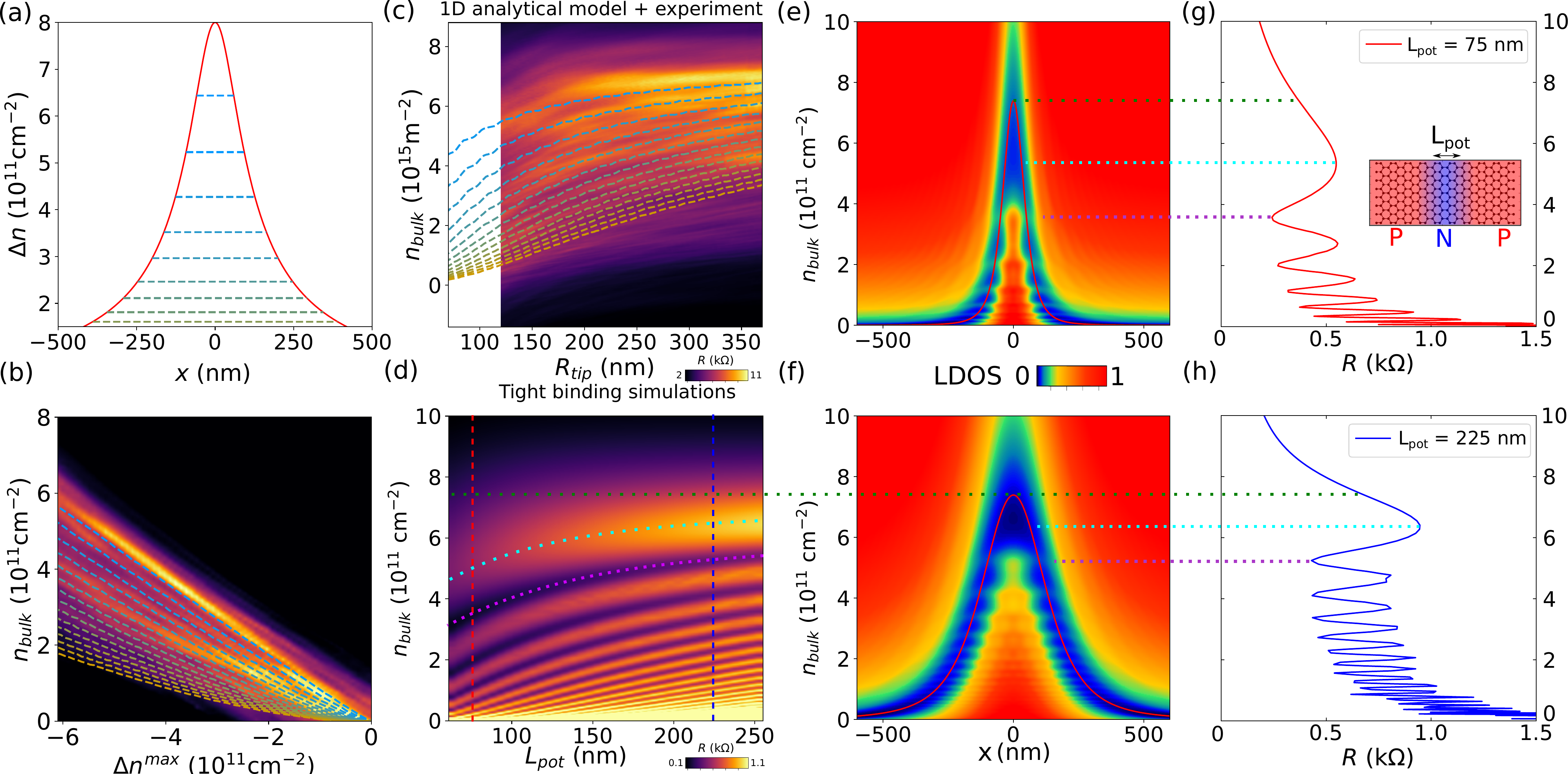}
\caption{\label{fig3}\textbf{Modeling the experiment:} (a) Example of tip-induced density change, and corresponding resonant modes energies calculated from Eq.(1). 
(b) Position of the 10 first modes as a function of $\Delta n^{max}$ and $n_{bulk}$ , for $R_{tip}$ = 250 nm, calculated from Eq.(1) and superimposed with the experimental data of Fig.~\ref{fig1}d.
(c) \textit{1D} analytical model of the position of Fabry-Pérot modes calculated from Eq.(1), with $\Delta n^{max}$ = $7,5.10^{15} \rm m^{-2}$. 
Dashed lines represent the bulk density corresponding to the 15 first resonances as a function of the potential extension $R_{tip}$, superimposed with the experimental data of Fig.~\ref{fig2}d. 
(d) Tight-binding model: simulated resistance as a function of the potential extension $L_{pot}$ and bulk density $n_{bulk}$, 
calculated for a graphene ribbon. (e) LDOS as a function of energy and \textit{x} position along the ribbon, showing the different resonances due to FP interferences, for a potential
extension $L_{pot}$ = 70 nm. Red line indicates the zero-density position. (f) Same calculation for $L_{pot}$ = 250 nm. (g) Calculated resistance as a function of $n_{bulk}$ for the same potential as in (e). 
Inset: schematics of the tight-binding system. (h) Calculated resistance as a function of $n_{bulk}$ for the same potential as in (f). 
}
\end{figure*}

To study the influence of this potential extension on the Fabry-P\'erot oscillations, we place the polarized tip on top of the constriction center (point C in Fig.~\ref{fig2}c),
and record the resistance as a function of $V_{bg}$, for different tip-to-graphene distance $d_{tip}$. As $d_{tip}$ is increased, we decrease $V_{tip}$ (towards more negative
values) to keep a constant value of $\Delta n^{max}$, and vary only the smoothness of the p-n junctions through $R_{tip}$. 
The resulting resistance map is plotted in Fig.~\ref{fig2}d and constitutes the main result of this study, together with its detailed theoretical analysis.
Figures \ref{fig2}e and \ref{fig2}f show the device resistance as a function of $V_{bg}$, for two extreme values of $R_{tip}$ in Fig.~\ref{fig2}d.
These two plots highlight two main features already visible in Fig.~\ref{fig2}d, \textit{i.e.}:\\
(i) The maximum value of the resistance increases with increasing  $R_{tip}$ as well as the density for which this maximum is reached.\\
(ii) The contrast of the Fabry-P\'erot interference evolves in a different way for the lower energy modes observed at low $V_{bg}$ 
(they decrease in amplitude) and the higher energy ones, whose amplitude increases with  $R_{tip}$.

In order to understand these observations, we analyze the problem with two different approaches. The first one is analytic: we use the potential landscape 
evaluated from Fig.~\ref{fig2}a and \ref{fig2}b, and follow the approach proposed in Ref.\cite{Drienovsky-2014} . 
We consider the tip potential as varying only along x-axis, and evaluate the position of the expected resonances from the simple equal phase condition: 
\begin{equation}
 2 \int_{-L_p}^{L_p} k(x) dx = 2 p \pi
\end{equation}
where $L_p$ is the position of zero charge density along x-axis which depends on the bulk density $n_{bulk}$, \textit{p} is a positive integer, and $k(x)$ is the position-dependent 
wave-vector evaluated from $n(x)$ provided that $k(x) = \sqrt{\pi n(x)}$.
Fig.~\ref{fig3}a shows a typical tip-induced density change and the position of the first resonant modes.
In Fig.~\ref{fig3}b, we calculate the expected position of the 10 first resonant modes for a tip potential extension of 250 nm as a function of $n_{bulk}$ and $\Delta n^{max}$, 
and report them as dashed lines on top of the experimental data of Fig.~\ref{fig1}d (where we have used the backgate and tip lever-arm parameters to 
convert the $V_{tip}$ and $V_{bg}$ axis into carrier densities). 
There is a good qualitative correspondence between the evolution of the different modes and the experimental fringes, 
reinforcing the interpretation of their origin as Fabry-P\'erot resonances inside the tip-induced n-p-n region.
Using $n(x)$ measured for the different couples of $d_{tip}$ and $V_{tip}$, displayed in Fig.~\ref{fig2}b, we also plot in Fig.~\ref{fig3}c the evolution of 
the first 15 modes in the ($n_{bulk}$,$R_{tip}$) plane, and find that they fall nicely on top of the experimental data of Fig.~\ref{fig2}d, rescaling 
the vertical axis $V_{bg}$ to a density and the horizontal axes $d_{tip}$ and $V_{tip}$ to the tip potential extension $R_{tip}$.

To go one step further in the understanding of the experimental fringes, we perform tight-binding simulations, using a home-made recursive Green functions code \cite{Nguyen-2010}.
We study a simple graphene ribbon, to which we apply a potential of variable extension $L_{pot}$ along transport direction,
(see inset of Fig.~\ref{fig3}g), with a smoothness governed by the exponent $\sigma$:
\begin{equation}
 V(x) = \frac{V_{max}}{1+(x/L_{pot})^\sigma }
\end{equation}
The ribbon width is fixed to 800~nm to avoid undesirable effects of transverse quantization (Fabry-P\'erot resonances are insensitive to the ribbon width).
We first consider a Lorentzian potential with $\sigma$ = 2, and calculate the ribbon resistance as a function of bulk density $n_{bulk}$ 
(\textit{i.e.} the charge carrier density in the p region) and potential extension $L_{pot}$, while keeping a fixed value 
of $\Delta n^{max}$ = $7,5.10^{11} \rm cm^{-2}$. The result is plotted in Fig.~\ref{fig3}d. 
First of all, it should be noted that the first resistance maximum (cyan dotted line on Fig.~\ref{fig3}d) is not obtained for $n_{bulk}$ = $-\Delta n^{max}$ (green dotted line).
This is clarified in Fig.~\ref{fig3}e, where we plot the local density of states (LDOS) in the graphene ribbon integrated over the transverse direction, as a function of $n_{bluk}$, aside
with the resistance as a function of bulk density (Fig.~\ref{fig3}g), for a potential extension $L_{pot}$ = 75 nm.
There is indeed a clear offset between the bulk density corresponding to the maximum of the tip-induced potential in Fig.~\ref{fig3}e 
(indicated by a green dotted line) and the resistance maximum in Fig.~\ref{fig3}g (indicated by a cyan dotted line).
The resistance maximum is rather reached for a density $n_{bulk}$ yielding a minimum LDOS at the barrier center.

Figures \ref{fig3}f and \ref{fig3}h present the same analysis for a larger potential extension.
In this case, the bulk densities corresponding to $\Delta  n^{max}$ (in green) and to the minimum LDOS (in cyan) are closer to each other, but still do not match.
The respective evolution of these two densities with the potential extension can be followed Fig.~\ref{fig3}d, as the spacing between the green and cyan dotted lines, 
and is in good agreement with the evolution of the resistance maximum observed in the experiment, as visible in figures \ref{fig2}d and \ref{fig3}c.

A second interesting feature well captured by this toy model is the evolution of the first resonant mode energy, 
visible as the first resistance minimum indicated by purple dashed lines in Figs.\ref{fig3}e-h, 
which follows roughly the same average evolution as the resistance maximum.
This first Fabry-P\'erot mode energy is reminiscent of the confinement energy due to the potential well created by the tip. 
In quantum mechanics, a famous textbook problem consists in finding the zero-point energy of a ``particle-in-a-box'', i.e. trapped in an 
infinite square potential of length $L$. The zero-point energy in the latter case emerges as a consequence of Heisenberg uncertainty principle, 
and increases with decreasing $L$ (as $L^{-2}$ for massive particles and $L^{-1}$ for massless Dirac fermions \cite{Cho-2011}). 
This distance of the first mode to the maximum of the tip potential is also clearly dependent on $R_{tip}$ in the experiment,
as visible ine Figs.\ref{fig2}d and \ref{fig3}c. It provides a nice illustration of this textbook problem, poorly explored in the 
case of Dirac fermions due to the inherent difficulty to confine them.

Discrepancies are however visible between results from this ideal ribbon model and experimental data. 
First of all, additional resonances are present in the experiment. They could result from the transverse quantization inherent to the narrow constriction, intentionally
suppressed in the tight-binding model by simulating a wide ribbon. These additional resonances could also arise from disorder, and the finite distance between the contacts 
and the constriction, that could lead to other Fabry-P\'erot cavities.
Secondly, the high resistance at low bulk density in the model is not present in the experiment. This can easily be understood as due to the experimentally measured 
finite resistance at the Dirac point, inherent to residual electron-hole puddles at low densities, whereas the tight-binding calculation in a homogeneous graphene ribbon
predicts a much larger resistance of the bulk (and leads) close to the Dirac point.
Both effects prevent the direct quantitative comparison of the interferences contrast in the experiment and the model presented in Fig.~\ref{fig3}.

\begin{figure}[h!]
\includegraphics[width = 0.6\linewidth]{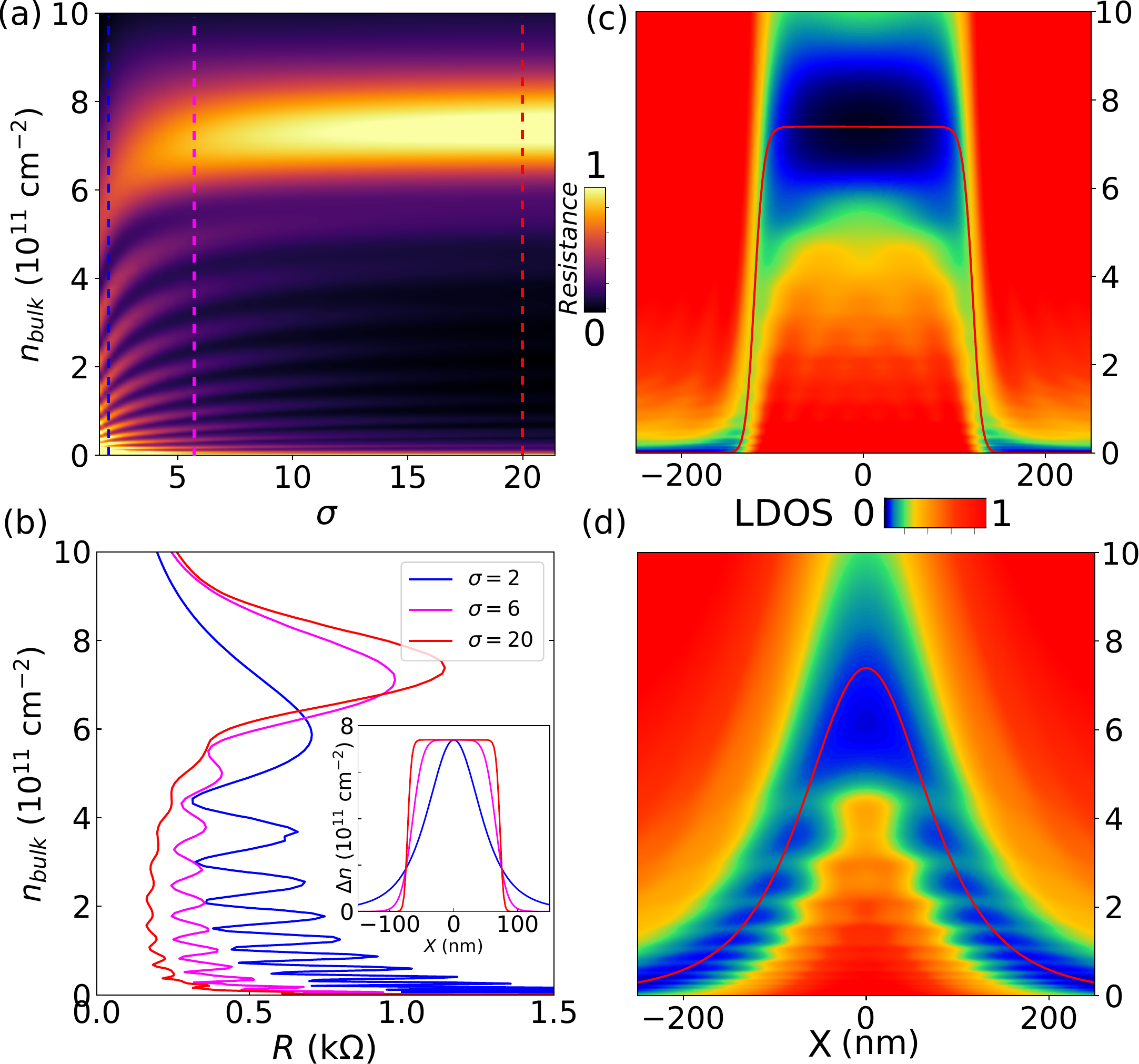}
\caption{\label{fig4}\textbf{Role of potential barriers smoothness:} (a) Tight-binding simulation of the Fabry-P\'erot interference fringes evolution: resistance of a ribbon 
as a function of $n_{bulk}$ and potential smoothness $\sigma$, with fixed $\Delta n^{max}$ = $7,5.10^{11} \rm cm^{-2}$ and extension $L_{pot}$ = 100 nm.
(b) Profiles extracted from the dashed lines in Fig.~\ref{fig4}a, corresponding to different decay exponent $\sigma$ from Lorentzian profile (blue, $\sigma = 2$) to abrupt step (red, $\sigma = 20$).
(c) Calculated LDOS for  $\sigma = 20$ corresponding to the red resistance curve in Fig.~\ref{fig4}b. (d) Calculated LDOS for  $\sigma = 2$,
corresponding to the blue resistance curve in Fig.~\ref{fig4}b.
}
\end{figure}

To better understand the influence of the tip potential extension, we perform additional tight-binding simulations and vary the potential steepness by changing
the decay exponent $\sigma$ in equation (2).
We first calculate the resistance of the ribbon as a function of bulk density and decay exponent, and plot the result in Fig.~\ref{fig4}a.
For three different decay exponents ($\sigma$ = 2,6,20) we extract the resistance as a function of $n_{bulk}$ and plot the result in Fig.~\ref{fig4}b.
These two figures evidence that the potential smoothness is a key ingredient, that governs the Fabry-P\'erot interference contrast.
Indeed, the relativistic nature of graphene charge carriers makes sharp potential barriers highly transparent due to Klein tunneling. 
As a consequence, the Fabry-P\'erot resonances in the LDOS are rather large and overlap (see Fig.~\ref{fig4}c), owing to their hybridization with the Dirac continuum of the bulk. 
This weak confinement yields poorly contrasted Fabry-P\'erot oscillations in the total resistance (red curve Fig.~\ref{fig4}b).
In contrast, a smooth p-n junction (on the Fermi wavelength scale) is a poor Dirac fermions transmitter, so that two facing smooth p-n junctions can 
be used to confine Dirac fermions in a more efficient way. This can be seen Fig.~\ref{fig4}d, where the LDOS in the case of a smooth n-p-n junction is plotted,
and exhibits well defined resonant modes, giving rise to pronounced Fabry-P\'erot oscillations in the resistance (blue curve Fig.~\ref{fig4}b).

The confinement of Dirac fermions in p-n nano-islands and the resulting LDOS resonances have recently been explored in a set of beautiful scanning tunneling 
microscopy experiments \cite{Gutierrez-2016,Zhao-2015,Lee-2016,Jiang-2017}. 
Although our SGM experiment does not give direct access to the LDOS, it allows to probe transport through such islands and
reveals the strength of LDOS resonances through Fabry-P\'erot oscillations in the device resistance.
Tight-binding simulations explicitly confirm that the interference contrast is related to the LDOS resonances strength, themselves governed by the p-n junction smoothness, 
which can be easily tuned in SGM, as demonstrated here.

In conclusion, we defined a n-p-n junction in a high mobility graphene sample using the polarized tip of a scanning gate microscope.
Oscillating patterns are observed in transport through the n-p-n junction that can be attributed to Fabry-P\'erot interferences. 
By simultaneously varying the tip-to-graphene distance and tip voltage, one can control and characterize the p-n junctions smoothness. 
In turn, this allowed to show that smoother p-n junctions induce a larger contrast of the interference fringes.
Using tight-binding simulations, we studied the influence of the p-n junctions smoothness on the LDOS resonances, resulting from the quasi-confinement of Dirac
fermions within the tip-induced potential. These LDOS resonances amplitude can be explicitly linked to the visibility of the Fabry-P\'erot oscillations.
In the quest towards ever reduced graphene devices size, gates are often placed as close as possible to the graphene plane. 
The present study recalls that the gate dielectric thickness governs the p-n junction smoothness, which strongly influences the visibility of interferences.
It then governs the efficiency of devices based on electron-optics concepts.
This underlines that these distances have to be cleverly adjusted in the conception of relativistic electron optics devices.

\vspace{1cm}

\section*{Acknowledgments}

The present research was funded by the F\'ed\'eration Wallonie-Bruxelles through the ARC Grant on 3D nanoarchitecturing of 2D crystals (No. 16/21-077) 
and from the European Union's Horizon 2020 Research and Innovation program (No. 696656).
B.B. (research assistant), N.M. (FRIA fellowship), B.H. (research associate), V.-H.N. and J.-C.C. (PDR No. T.1077.15 and ERA-Net No. R.50.07.18.F) 
acknowledge financial support from the F.R.S.-FNRS of Belgium.
Support by the Helmholtz Nanoelectronic Facility (HNF), the EU ITN SPINOGRAPH and the DFG (SPP-1459) is gratefully acknowledged.
Growth of hexagonal boron nitride crystals was supported by the Elemental Strategy Initiative conducted by the MEXT, Japan and JSPS KAKENHI Grant Numbers JP26248061, JP15K21722 and JP25106006.
B.B aknowledges the use of Kwant \cite{Groth-2014} used to guide the experiment and cross-check tight-binding simulations.

\section*{References}
\newpage

\begin{flushleft}

\includegraphics[page=1,scale=0.85]{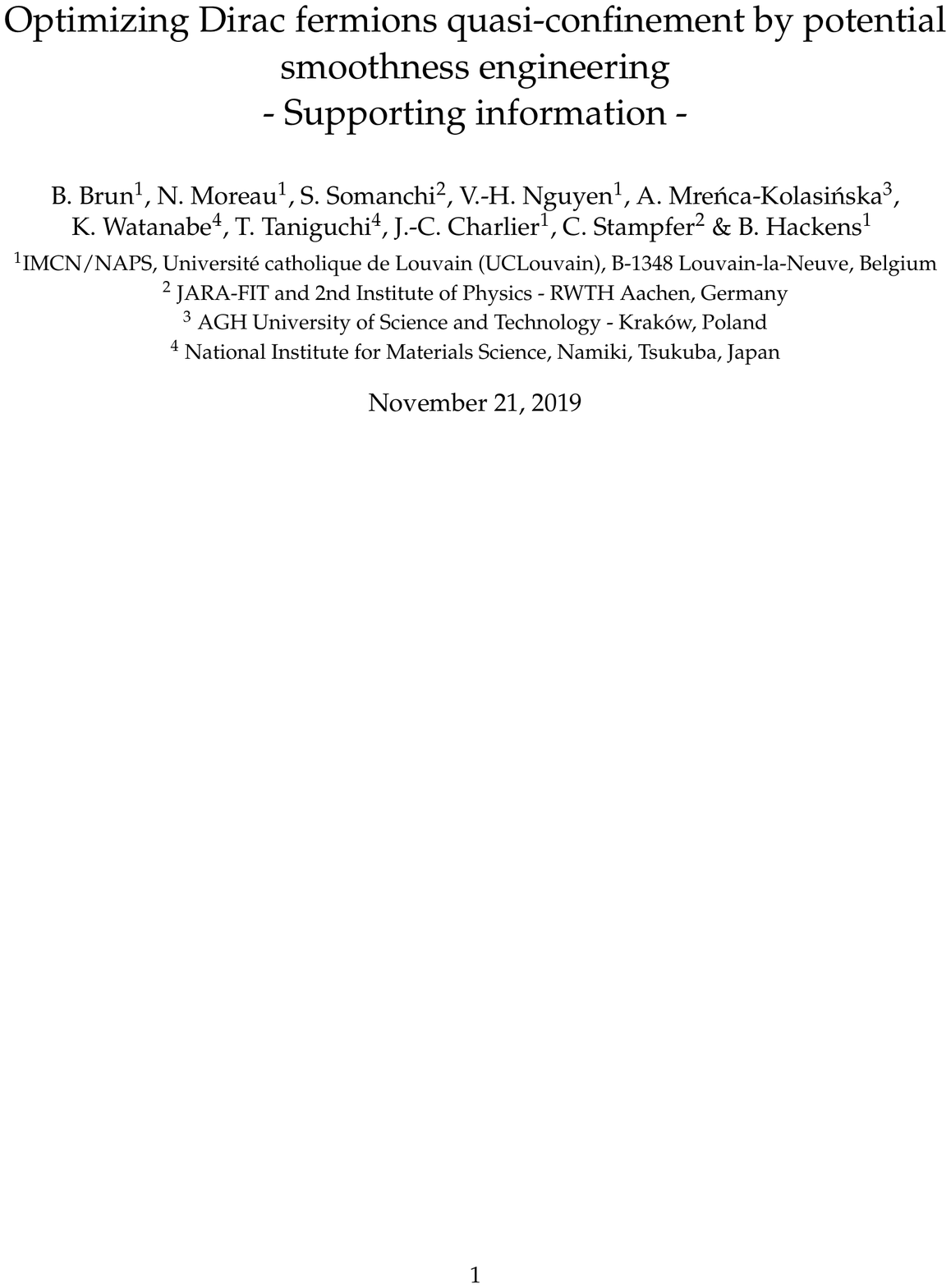} 
\includegraphics[page=2,scale=0.85]{supporting.pdf} 
\includegraphics[page=3,scale=0.85]{supporting.pdf} 
\includegraphics[page=4,scale=0.85]{supporting.pdf} 
\includegraphics[page=5,scale=0.85]{supporting.pdf} 
\includegraphics[page=6,scale=0.85]{supporting.pdf} 
 
\end{flushleft}

\end{document}